\documentclass[letterpaper, aps, prl, twocolumn, superscriptaddress, showpacs, nofootinbib]{revtex4}
\pdfoutput=1

\usepackage{graphicx}
\usepackage{bm}
\usepackage{amssymb}
\usepackage{amsmath}
\usepackage{amsfonts}
\usepackage{dcolumn}
\usepackage{natbib}
\usepackage{subfigure}
\usepackage{float}
\usepackage{siunitx}
\usepackage[colorlinks,citecolor=blue,urlcolor=blue]{hyperref}
\pacs{04.25.D-, 04.25.dg, 04.30.Db}

\usepackage[usenames, dvipsnames]{xcolor}

\usepackage[utf8]{inputenc}

\newcommand{\raiseentry}[1]{\smash{\raise 0.7 em \hbox{#1}}}

\def\apj{Astrophys. J.}
\def\apjl{Astrophys. J. Lett.}
\def\apjs{Astrophys. J. Supp. Ser.}

\def\mnras{Mon. Not. Roy. Astron. Soc. }

\def\prl{Phys. Rev. Lett.}
\def\prd{Phys. Rev. D.}

\def\cqg{Class. Quantum Grav.}

\newenvironment{equationarray*}
{\arraycolsep 0.14 em
\begin{eqnarray*}}
{\end{eqnarray*}}

\begin{document}

\title{Gravitational Waves from Binary Black Hole Mergers Inside of Stars}

\author{Joseph M. Fedrow}
\email{jfedrow@yukawa.kyoto-u.ac.jp}
\affiliation{Center for Gravitational Physics and International Research Unit of Advanced Future Studies, Yukawa Institute for Theoretical Physics, Kyoto University, Kyoto, Japan}

\author{Christian D. Ott}
\affiliation{Center for Gravitational Physics and International Research Unit of Advanced Future Studies, Yukawa Institute for Theoretical Physics, Kyoto University, Kyoto, Japan}
\affiliation{TAPIR, Walter Burke Institute for Theoretical Physics, California Institute of Technology, Pasadena, CA, USA}

\author{Ulrich Sperhake}
\affiliation{Department of Applied Mathematics and Theoretical Physics, Centre for Mathematical Sciences, University of Cambridge, Cambridge, United Kingdom}
\affiliation{TAPIR, Walter Burke Institute for Theoretical Physics, California Institute of Technology, Pasadena, CA, USA}

\author{Jonathan~Blackman}
\affiliation{TAPIR, Walter Burke Institute for Theoretical Physics, California Institute of Technology, Pasadena, CA, USA}

\author{Roland Haas}
\affiliation{National Center for Supercomputing Applications, University of Illinois at Urbana-Champaign, 1205 W Clark St, Urbana, IL 61801, USA}

\author{Christian Reisswig}
\affiliation{TAPIR, Walter Burke Institute for Theoretical Physics, California Institute of Technology, Pasadena, CA, USA}

\author{Antonio De Felice}
\affiliation{Center for Gravitational Physics and International Research Unit of Advanced Future Studies, Yukawa Institute for Theoretical Physics, Kyoto University, Kyoto, Japan}

\date{\today}

\begin{abstract}
We present results from a controlled numerical experiment
investigating the effect of stellar density gas on the coalescence of
binary black holes (BBHs) and the resulting gravitational waves
(GWs). This investigation is motivated by the proposed stellar core
fragmentation scenario for BBH formation and the associated
possibility of an electromagnetic counterpart to a BBH GW event. We
employ full numerical relativity coupled with general-relativistic
hydrodynamics and set up a $30 + 30 M_\odot$ BBH (motivated by
GW150914) inside gas with realistic stellar densities. Our results
show that at densities $\rho \gtrsim 10^6 - 10^7 \, \mathrm{g \,
  cm}^{-3}$ dynamical friction between the BHs and gas changes the
coalescence dynamics and the GW signal in an unmistakable way. We show
that for GW150914, LIGO observations appear to rule out BBH
coalescence inside stellar gas of $\rho \gtrsim 10^7 \,
\mathrm{g\,cm}^{-3}$. Typical densities in the collapsing cores of
massive stars are in excess of this density. This excludes the
fragmentation scenario for the formation of GW150914.
\end{abstract}

\maketitle

\noindent \textbf{\textit{Introduction.}} With the recent detection of
the first gravitational wave (GW) events by LIGO
\cite{abbottphysrevx}, the era of GW Astronomy has begun. An extensive
multi-wavelength network of astronomical observatories is following up
each candidate GW event with the hope of catching an electromagnetic
(EM) counterpart. This is very well motivated for GWs from neutron
star (NS) mergers (e.g., \cite{metzger:12}), but for observed GWs from
the merger of stellar-mass binary black holes (BBHs), no EM
counterpart is expected (e.g., \cite{LIGO_bbh_counterpart}).

However, the first observed BBH GW event, GW150914
\cite{abbott2016observation}, was possibly connected with a 
$\gamma$-ray event seen by the Fermi satellite
\cite{connaughton:16} (though note it was not observed by
other $\gamma$-ray satellites
\cite{abbott2016localization,hurley2016interplanetary}). If directly
related, this would be a totally unexpected EM counterpart to what was
believed to be a BBH merger in pure vacuum.

To explain such an EM counterpart, Loeb \cite{loeb2016electromagnetic}
proposed that the coalescing BHs formed via dynamical fragmentation in
a very massive star undergoing gravitational collapse. This scenario
is tentatively supported by the simulations of
\cite{reisswig2013formation}, who found BBH formation by dynamical
fragmentation in pair-unstable supermassive primordial stars. The
result of Loeb's scenario would be a BBH system embedded in
high-density stellar gas whose coalescence could drive outflows giving
rise to the $\gamma$-ray transient observed by Fermi.

There are arguments from stellar evolution
\cite{woosley2016progenitor} suggesting it may be difficult to obtain
collapsing stellar cores permitting dynamical
fragmentation. However, this possibility is not conclusively ruled out
by theory.

In this \emph{Letter}, we consider the scenario in which a BBH was
formed inside a collapsing massive star and conduct the first
numerical relativity simulations of BBH mergers in the presence of gas
with densities comparable to those in the cores of collapsing massive
stars. The results of our simulations show that the GWs observed from
GW150914 \emph{are inconsistent} with this event having taken place
inside a collapsing massive star, ruling out the dynamical fragmentation
scenario.

\textbf{\textit{Methods and Initial Data.}} We employ the open-source
\texttt{Einstein Toolkit} and evolve Einstein's equations in the BSSN
formalism \cite{shibata:95,baumgarte:99} with fourth-order finite
differences and adaptive mesh refinement (AMR). We include
general-relativistic (GR) hydrodynamics in the finite-volume approach with
piecewise parabolic reconstruction at cell interfaces and the Marquina
flux formula for intercell fluxes \cite{aloy:99}. Inside the BH
apparent horizons, we correct unphysical states using the methods
detailed in \cite{faber:07,farris:10}. Spacetime and hydrodynamics
evolution are coupled in a fourth-order Runge-Kutta integrator.
  
For generality, we describe our setup in $G = c = 1$ units and measure
quantities in terms of the ADM mass $M$. We employ BBH puncture
initial data and carry out a vacuum simulation (model G0) and four
simulations in which we embed the BBH system in gas of constant
density $\rho_0 = \{10^{-10}, 10^{-9}, 10^{-8}, 10^{-7}\} M^{-2}$
initially at rest, labeled G1--G4 in the order shown. We use
\texttt{TwoPunctures} \cite{ansorg:04,loeffler:06} to solve for
constraint satisfying quasi-circular initial data, taking into account
the gas, and placing the two equal-mass, non-spinning punctures at a
coordinate separation of $11.6\,M$. In the vacuum case, this
corresponds to 8 orbits to merger. We employ a $\Gamma$-law equation
of state $P = (\Gamma - 1) \rho \epsilon$ for the gas. We set $\Gamma
= 4/3$ and obtain the initial $\epsilon$ by assuming a gas dominated
by relativistic degenerate electrons (e.g., \cite{shapteu:83}). We
smoothly reduce $\rho$ to an atmosphere value of $10^{-16}\,M^{-2}$
outside of $80\,M$ by applying $X(R) = 0.5 [ 1 + \tanh( [R - 80 M] /
  15 M ) ]$.

We employ 7 levels of AMR with the outer boundary placed at $320\,M$.
The punctures are covered with a finest grid of $\Delta x =
0.0195\,M$, which corresponds to approximately 45 grid points across
each apparent horizon after an initial gauge adjustment. The fine grid
has a linear extent of $3\,M$ to provide high resolution for the gas
dynamics near the horizons. It is embedded in 5 coarser
AMR levels tracking the punctures' orbital motion. The outermost two
levels are fixed.  We extract GWs at $R = 120\,M$ where $\Delta x =
0.625\,M$ using the Newman-Penrose $\Psi_4$
formalism~\cite{loeffler:12,newman:62}. We obtain the GW strain via
fixed-frequency integration~\cite{reisswig:11}.
  
Rescaled to a BBH mass of $M = 60\,M_\odot$ for comparison with
GW150914, each puncture has an approximate initial mass of
$30\,M_\odot$, the initial separation is $1030\,\mathrm{km}$, with gas
densities $\rho_0 = 1.72 \times
\{10^4,10^5,10^6,10^7\}\,\mathrm{g\,cm}^{-3}$. The typical central
density in a presupernova star is
$10^{9}-10^{10}\,\mathrm{g\,cm}^{-3}$. At a radius of
$1000\,\mathrm{km}$ it is in the range $10^{7} -
10^{9}\,\mathrm{g\,cm}^{-3}$, depending on stellar mass (e.g.,
\cite{whw:02}). We choose $10^7\,\mathrm{g\,cm}^{-3}$ as the highest
simulated density since it is a reasonable and arguably low value for
the density of outer stellar core material left surrounding the BBH
formed in dynamical fragmentation. With the above choices, the total
gas mass on the computational grid is
$\sim$$13.8\,[M/(60\,M_\odot)][\rho_0 /
  (1.72\times10^7\,\mathrm{g\,cm^{-3}})]\,M_\odot$.

We provide a convergence study and analysis details
in the \emph{Supplemental Material} to this Letter.

\smallskip

\begin{figure}[h]
  \includegraphics[width=\linewidth]{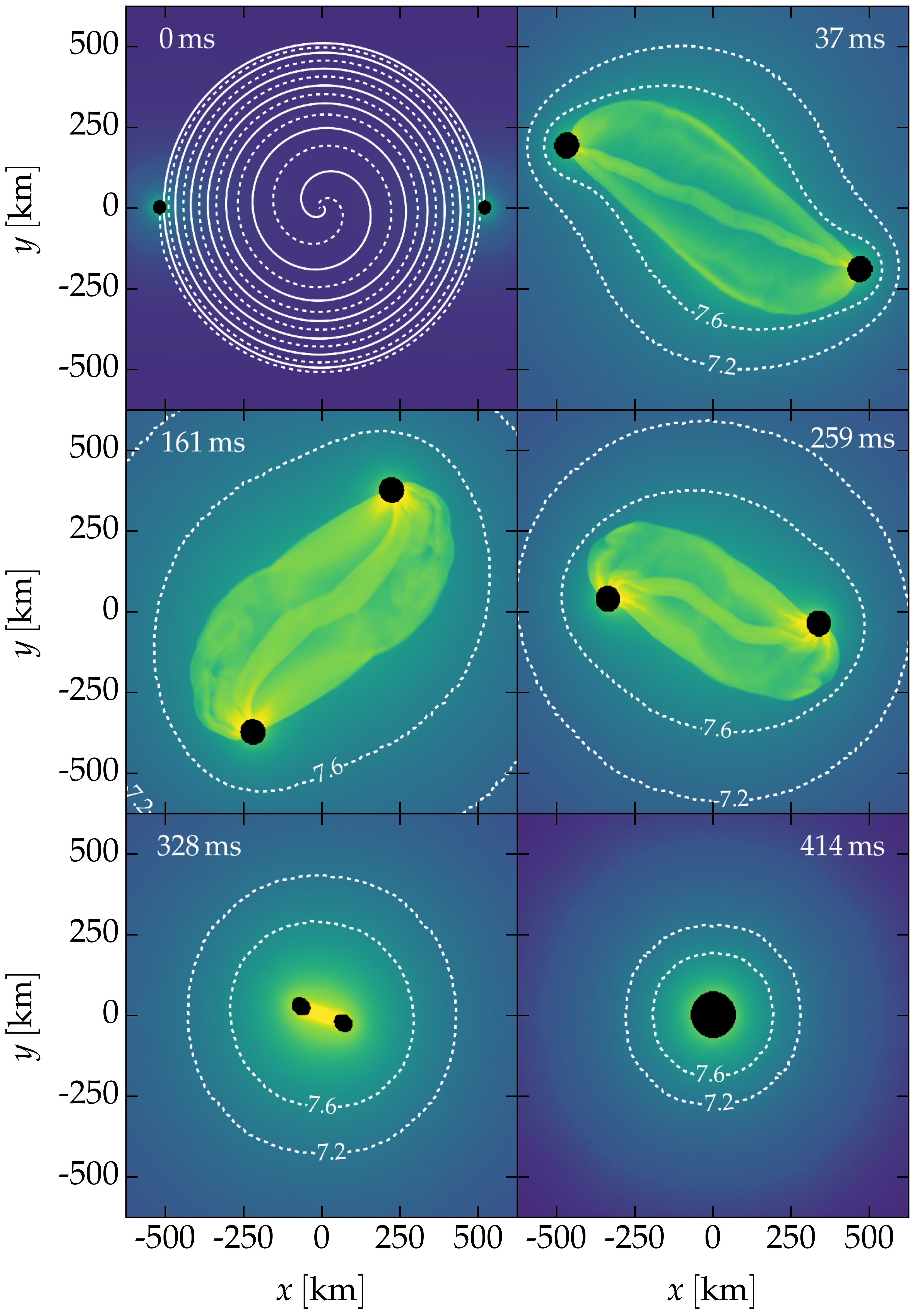}
  \vspace*{-0.08cm}
  
  \hspace*{1.05cm}\includegraphics[width=0.7\linewidth]{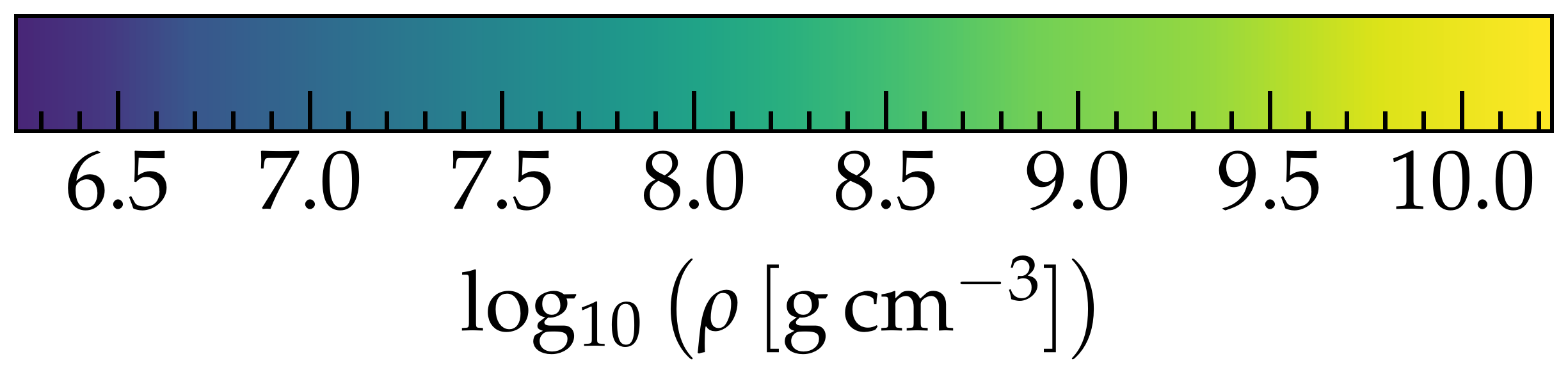}
  
\caption{\small{BBH inspiral evolution and orbital plane density
    slices of the G3 model ($\rho_0 \sim 10^{6} \, \mathrm{g
      \,cm^{-3}}$). The top-left frame shows the orbital tracks
    followed by the BBH in the subsequent frames. The top-right frame
    shows the emergence of a high-density gas bar due to gravitational
    focusing of gas between the BHs. We choose representative
    isocontours at $\rho = 10^{7.2} \, \mathrm{and} \, 10^{7.6} \,
    \mathrm{g \, cm^{-3}}$ to visualize the formation of ellipsoidal
    density structures surrounding the BBH. Initially, the orbital
    separation $a$ decreases slowly. Gas accumulates around the BBH
    pushing the isocontours to larger radii (center-left frame). Once
    $a$ is decreasing rapidly, the contours contract and circularize
    (center-right frame, bottom-left frame).  The bottom-right frame
    shows the final merged BH evolving toward steady-state Bondi-Hoyle
    accretion.}}
  \label{fig:colormap}
\end{figure}

\textbf{\textit{Dynamics.}} In Fig.~\ref{fig:colormap}, we show
orbital-plane snapshots of the rest-mass density at various times in
model G3's coalescence. In cgs units and for the $M = 60\,M_\odot$
case, its merger time is $\sim390\,\mathrm{ms}$ (we define merger time
based on the peak amplitude of the $(2,2)$ GW mode). That is
$\sim142\,\mathrm{ms}$ faster than the pure-vacuum case G0.

The density colormaps in Fig.~\ref{fig:colormap} reveal that soon
after the start of the simulation, an ellipsoidal high-density
structure surrounds the BHs. The central high-density band visually
connecting the BHs is due to the gravitational focusing of gas into
this region, where acceleration toward one BH is partially cancelled
by the other. This feature was also observed in BBH mergers in very
low-density gas (e.g., \cite{bode:12,farris:10} and references
therein).

The ellipsoidal stucture surrounding the BBH in
Fig.~\ref{fig:colormap} forms because each BH accelerates the
surrounding gas, dragging it along in its gravitational wake. The
associated drag force, closely related to dynamical friction (e.g.,
\cite{escala:04, chandra:43,
  ostriker:99,dai2016energetic,barausse:07,barausse:14}), converts
orbital energy into kinetic energy and internal energy of the gas
(through compression and shocks). This process is what rapidly robs
the BBH of its orbital energy and angular momentum. It leads to an
accelerated decline of the orbital separation and an earlier merger
compared to the vacuum case G0.

The BHs accrete gas during coalescence, but even in the high-density
G4 case, the total mass accreted by each BH is only $\sim$4\% of its
initial mass. The effect of the gradually changing mass on the
coalescence is much smaller than that of dynamical friction.

\begin{table}[t]
  \caption{Model Summary. $\rho_0$ is the initial gas density,
    $t_\mathrm{merge}$ the merger time, and
    $\mathcal{M}_\mathrm{ZDHP}$ and $\mathcal{M}_{150914}$ are the GW
    mismatches with the vacuum waveform for Advanced LIGO design noise
    and noise at the time of GW150914, respectively. For GW150914, a
    mismatch $\mathcal{M} \gtrsim 0.0017$ becomes noticable.}
\label{tab:parameters}
\begin{ruledtabular}
\begin{tabular}{lccll} 
Model 
&\multicolumn{1}{c}{$\rho_0\, (M / 60 M_{\odot})^{-2}$}
&\multicolumn{1}{c}{$t_\mathrm{merge}$}
&\multicolumn{1}{c}{$\mathcal{M}_\mathrm{ZDHP}$}
&\multicolumn{1}{c}{$\mathcal{M}_{150914}$}  \\
&\multicolumn{1}{c}{$[\textrm{g cm}^{-3}]$}
&\multicolumn{1}{c}{[ms]} & \\[0.1em]
\hline
G0 Vacuum & 0             & 510    & 0     & 0 \\
G1     & $1.72\times10^4$ & 508    & $8 \times 10^{-5}$ & $3 \times 10^{-5}$     \\
G2     & $1.72\times10^5$ & 490    & $0.0058$     & $0.0016$ \\
G3     & $1.72\times10^6$ & 369    & $0.1882$  	  & $0.0665$ \\
G4     & $1.72\times10^7$ & 186    & $0.3718$     & $0.2386$  \\
\end{tabular}
\end{ruledtabular}
\end{table}

\begin{figure}
  \includegraphics[width=\linewidth]{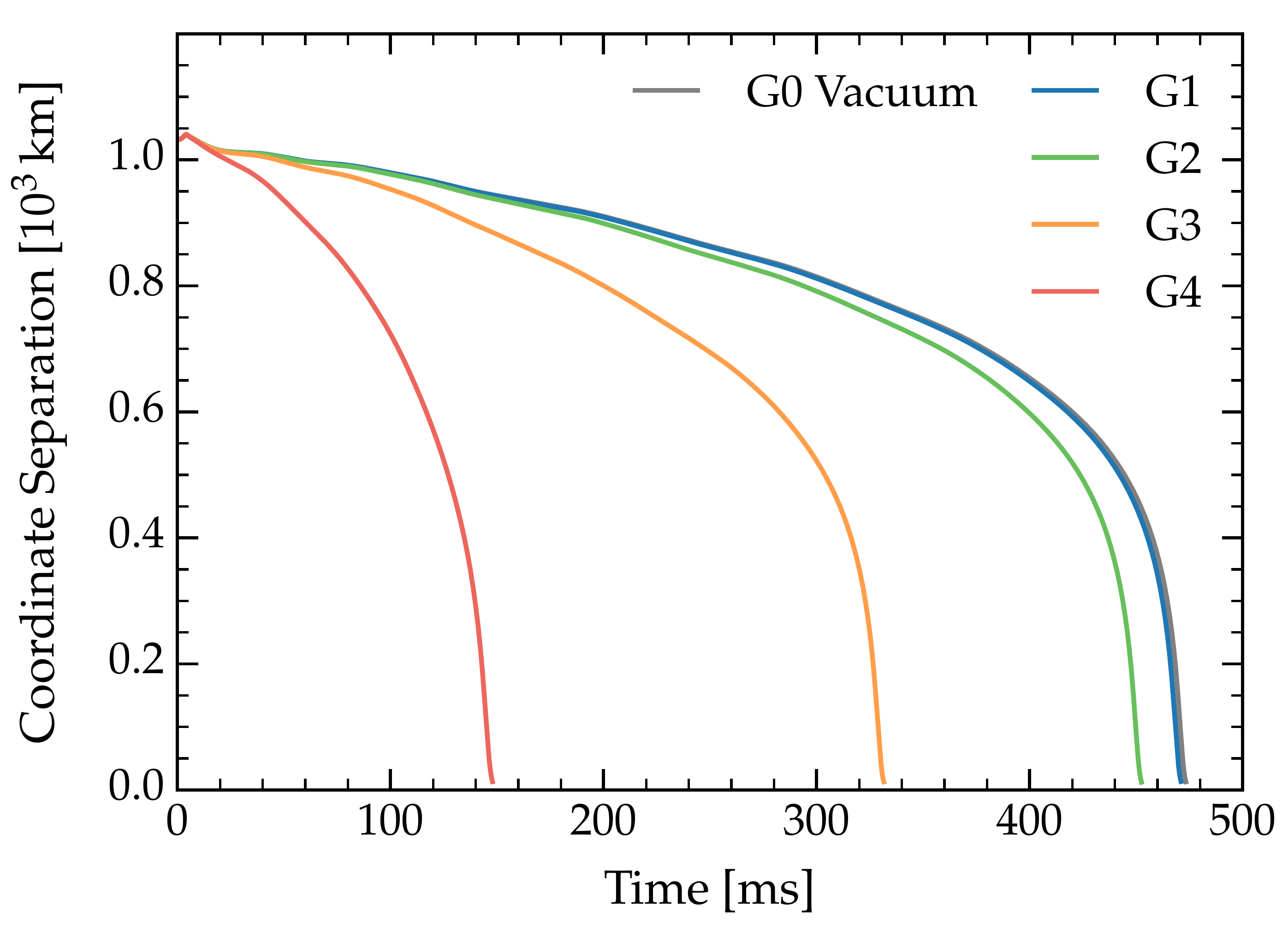}
  \vspace{-0.6cm}
  \caption{BBH coordinate separation $a$ as a function of time until
    common horizon formation. All simulations start from the same
    separation of $1030$ km (assuming a total BBH mass of $60 \,
    M_\odot$). As $\rho_0$ increases across models G1--G4, dynamical
    friction dissipates orbital energy resulting in earlier mergers.}
  \label{fig:coordsep_instfreq_from_h}
\end{figure}

\begin{figure}[t]
  \includegraphics[width=\linewidth]{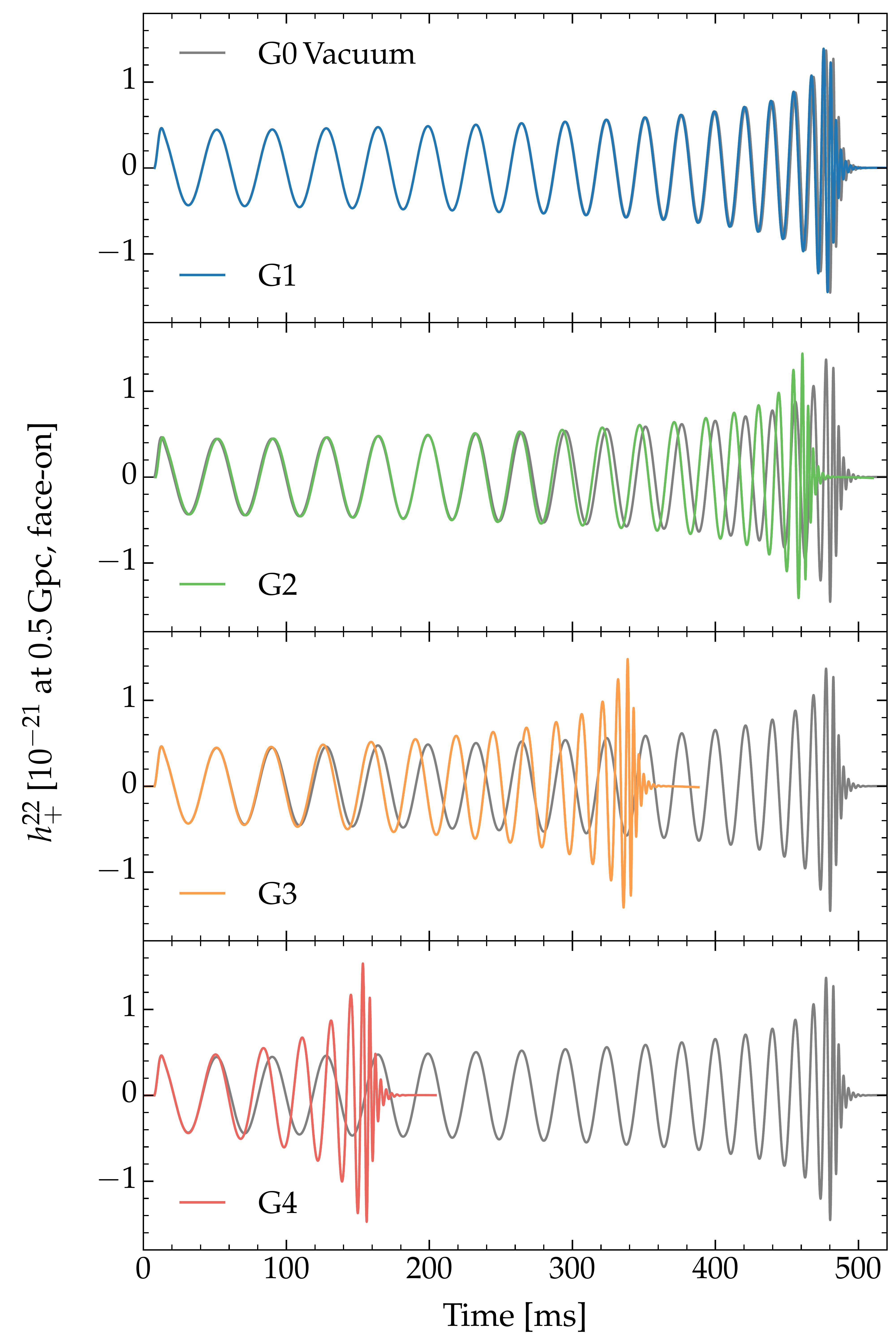}
  \vspace{-0.6cm}
  \caption{Real part of the $(2,2)$ GW strain, observed face-on from a
    distance of $0.5$ Gpc for a total BBH mass of $60 \, M_\odot$. We
    compare models G1--G4 with G0 vacuum plotted in gray in each
    panel. All GWs start with $f \sim 26\,\mathrm{Hz}$. Increasing
    density leads to faster chirps. The ringdown frequency is $\sim
    296 \, \mathrm{Hz}$ for G0 vacuum, decreasing by only $\sim10 \%$
    for G4.  }
  \label{fig:waveforms}
\end{figure}

In Tab.~\ref{tab:parameters}, we summarize key properties, including
the merger times, for all simulated models. The top panel of
Fig.~\ref{fig:coordsep_instfreq_from_h} shows the BBH coordinate
separation $a$ as a function of simulation time. With increasing gas
density, the merger is driven to earlier times. Model G1
($\rho_0\sim10^4\,\mathrm{g\,cm^{-3}})$ is only very mildly affected,
merging some $\Delta t \sim 2$ \,ms earlier than the vacuum case
G0. $\Delta t$ is $\sim 20\,\mathrm{ms}$, $\sim 142\,\mathrm{ms}$, and
$\sim 324\,\mathrm{ms}$, for models G2, G3, and G4, respectively,
which have $10$, $100$, and $1000$ times higher density than model
G1. The effect of the gas on the coalescence time is thus roughly
linear in density for the lower-density cases. This is qualitatively
reproduced by a simple Newtonian point-particle model including GW
($\partial a/\partial t \propto a^{-3}$) and dynamical friction
($\partial a/\partial t \propto a^{5/2}$) \cite{escala:04} terms for
orbital evolution. At high density, i.e.\ going from G3 to G4,
dynamical friction is so strong that it is no longer a linear
perturbation to the GW-dominated inspiral.  The point-particle model
shows that in G1--G3, the dynamical friction term is always
subdominant. In G4, it initially dominates over the GW term, but
quickly decreases in relevance as the orbit shrinks and GW-driven
evolution takes over.

\textbf{\textit{Gravitational Waves.}} In Fig.~\ref{fig:waveforms}, we
present $h^{22}_+$, the real part of the $l=2,m=2$ GW mode.  The low
density in model G1 has a negligible effect and its waveform is
essentially identical to vacuum GWs.
As the density increases from G1 to G4, merger occurs progressively
earlier. This leads to dramatic changes in the emitted GW train and
creates an unmistakable GW signature: (1) Since all models start at
the same separation, the initial GW frequency is $f_0
\sim$$26\,\mathrm{Hz}$ for all models. (2) Dynamical friction shortens
the inspiral, leading to a faster sweep (``chirp'') of the GWs through
frequency space. (3) The faster chirp is not due to a substantial
increase of the BBH mass. Hence, merger and ringdown GW emission is at
frequencies that change only mildly with $\rho_0$.  Model G0 has a
final BH mass of $\sim$$57.6\,M_\odot$, a dimensionless spin $a^\star
= 0.69$, and we find a ringdown GW frequency of
$\sim$$296\,\mathrm{Hz}$. The highest-density model G4 produces a
merged BH of $\sim$$64.8\,M_\odot$ and $a^\star = 0.65$, consistent
(see, e.g., \cite{berti:09}) with its ringdown GW frequency of
$\sim$$265\,\mathrm{Hz}$.

\textbf{\textit{Analysis and Observational Consequences.}}  We compute
the GW mismatch $\mathcal{M}(h_i,h_j)$ (see
\cite{damour:98,reisswig:11} and \emph{Supplemental Material}) for
each waveform G1--G4 with the G0 vacuum case. GW mismatch takes into
account the detector noise spectrum and we consider frequencies in the
interval $[26,3000]\,\mathrm{Hz}$. We employ Advanced LIGO design
noise \cite{LIGO-sens-2010} ($\mathcal{M}_\mathrm{ZDHP}$) and the
noise around GW150914 \cite{abbott2016observation}
($\mathcal{M}_{150914}$).  $\mathcal{M}$ is in $[0,1]$ and
$\mathcal{M} = 0$ means $h_i$ and $h_j$ are identical.  For an
observation with signal-to-noise ratio $\varrho$, an $\mathcal{M}
\gtrsim 1/\varrho^2$ leads to observational inconsistencies (see
\cite{lindblom:08} and \emph{Supplemental Material}). GW150914 was
observed with $\varrho \sim 24$, so $\mathcal{M} \gtrsim 0.0017$ will
become noticable.

We summarize $\mathcal{M}_\mathrm{ZDHP}$ and
$\mathcal{M}_\mathrm{150914}$ for all models in
Tab.~\ref{tab:parameters}. The results for
$\mathcal{M}_\mathrm{150914}$ show that for GW150914, densities
$\rho_0 \lesssim 10^{4}\,\mathrm{g\,cm}^{-3}$ (G1) are
indistiguishable from vacuum.  Model G2 ($\rho_0 \sim
10^5\,\mathrm{g\,cm}^{-3}$) is marginally distinguishable.  The
situation is very different for models G3 and G4 with
$\mathcal{M}_\mathrm{150914}$ $\sim$$0.07$ and $\sim$$0.24$,
respectively. These results show that stellar densities $\rho_0
\gtrsim 10^6\,\mathrm{g\, cm}^{-3}$ lead to highly significant
inconsistencies with vacuum.

An additional possibility is that the G1--G4 waveforms could have
lower mismatch with vacuum waveforms of BBHs with different
parameters. We explore this with a 7-dimensional numerical relativity
surrogate GW model~\cite{blackman:17,blackman:17b}, covering BBH mass
ratio $q$ (up to $q=2$) and six spin components (up to effective spin
$\chi_\mathrm{eff} = (M_1 a^*_1 + M_2 a^*_2) / M = 0.8$), assuming
zero eccentricity. We vary parameters to minimize
$\mathcal{M}_\mathrm{150914}$ and find $0.011$ and $0.061$, for model
G3 and G4, respectively.  For G3, the minimum
$\mathcal{M}_\mathrm{150914}$ is at $M = 70.6\,M_\odot$, $q \simeq
1.0$, and effective spin $\chi_\mathrm{eff} \simeq 0.17$.  For G4, we
find $M = 75.3\,M_\odot$, $q \simeq 1.6$, and $\chi_\mathrm{eff}
\simeq -0.47$. Even with the reduction in
$\mathcal{M}_\mathrm{150914}$, $\rho_0 \gtrsim
10^{7}\,\mathrm{g\,cm}^{-3}$ leads to observable differences with any
waveform covered by the surrogate model.

Having established that an equal mass, non-spinning BBH merger in
stellar-density gas with $\rho_0 \gtrsim 10^7\,\mathrm{g\,cm}^{-3}$ is
inconsistent with GW150914 and all BBH waveforms from our
  surrogate model, there remains the following crucial question: Are
there BBH parameter choices that could make a merger in gas appear
just like GW150914?

We argue that the answer is 'No': The observational BBH parameter
space encompasses total mass, mass ratio, eccentricity, and six spin
components. (i) BBHs of lower $M$ inspiral to higher frequencies and
have more cycles from $\sim$26~Hz to merger. Using the surrogate, we
find that $M = 43.7\,M_\odot$ extends the G0 case by
  $\sim$324\,ms, the difference in merger times between G0 and
  G4. However, its time-frequency evolution and ring-down frequency
($\sim$$400\,\mathrm{Hz}$) are substantially different from GW150914,
leading to large mismatch. (ii) Dynamical fragmentation in our
scenario leads to near-equal-mass fragments (e.g.,
\cite{reisswig2013formation}). We consider $q=2$ as
  an extreme limit. In the vacuum case, it extends the inspiral by
$\sim$$38\,\mathrm{ms}$ \cite{blackman:17b}, insufficient to
compensate for the gas effect. (iii) High BH spin causes ``orbital
hangup.'' The effect is largest for equal spins aligned with the
orbital angular momentum. Using our surrogate and the SpEC waveforms
\cite{SXSCatalog}, we find that for $a^* = 0.99$ ($a^* = 0.4$),
inspiral is prolonged by $177\,\mathrm{ms}$ ($71\,\mathrm{ms}$). The
effect is linear in $a^*$. To explore the effect of spin in the
stellar-density G4 case, we carry out a simulation with $a^* = 0.4$
for both BHs.  We find that merger is delayed by
$\sim$$17.2\,\mathrm{ms}$.  Extrapolating to $a^* = 0.99$ from the
vacuum case, spin could extend the G4 inspiral by at most
$\sim$$39\,\mathrm{ms}$. This is insufficient to mimic GW150914.  


\textbf{\textit{Discussion and Conclusions.}} Fragmentation of a
massive star's core into clumps that collapse further to NSs or BHs is
an interesting scenario for the formation of NS binaries and BBHs
(e.g.,
\cite{bonnell:95,davies:02,postnov:16,reisswig2013formation}). While
perhaps unlikely (e.g., \cite{fryer:04,woosley2016progenitor}), this
scenario has not previously been ruled out observationally. As
proposed by Loeb \cite{loeb2016electromagnetic}, it would endow a BBH
merger with the gas necessary to produce an EM counterpart. Dai~\emph{et
  al.}~\cite{dai2016energetic} suggested, but did not show, that
the gas surrounding the BBH could have observable consequences in the
emitted GWs.

We employed numerical relativity coupled with GR hydrodynamics for a
controlled experiment into the effects of stellar-density gas on BBH
mergers.  Scaled to a total system mass of $60\,M_\odot$ (consistent
with GW150914), our results show that dynamical friction between the
BHs and gas at stellar densities $\rho_0 \gtrsim$$10^6 -
10^7\,\mathrm{g\,cm}^{-3}$ profoundly affects the coalescence
dynamics, drastically shortening the time to merger. This modifies the
resulting GW signal in an unmistakable way, leading to differences
with vacuum waveforms that can be observed by LIGO.

Our analysis furthermore suggests that it is not
  possible to choose BBH parameters that would yield a waveform in
  stellar-density gas resembling GW150914\cite{ligosys:17}. Thus we
  conclude that it is highly unlikely that GW150914 was formed through
  dynamical fragmentation in a massive star and Loebs scenario
  \cite{loeb2016electromagnetic} is ruled out by the GW observation
  alone.  

Future work should address the limitations of our work: We
assumed the gas to be non-magnetized and initially at rest, but
angular momentum and magnetic fields can have dynamical impact. We
employed a constant density, but real stars have radially varying
density. Finally, we used a $\Gamma$-law equation of state, ignoring
microphysics such as electron capture, neutrinos, and nuclear
dissociation, which all may have effects on the gas dynamics. In our
analysis, we did not consider GW detector calibration uncertainties of
$\sim$$10\%$ \cite{abbott2016observation}. This should affect all
waveforms equally and is unlikely to alter our conclusions.

\smallskip

We provide waveforms and additional visualizations of our simulations
at \url{https://stellarcollapse.org/bbhgas}. We thank M.~Sasaki,
G.~Dom\`enech, K.~Kiuchi, M.~Shibata, K.~Ioka, T.~Tanaka, E.~Schnetter,
E.~Firing, T.~Bogdanovic, and N.~Deruelle for 
discussions. This research is partially supported by MEXT, IRU-AFS,
NSF grants ACI-1550514, CAREER PHY-1151197, and PHY-1404569, and
ERC-2014-CoG 646597, MSCA-RISE-2015 690904, and STFC ST/L000636/1.  We
used the \texttt{matplotlib} Python package \cite{hunter:07} for the
figures. The simulations were performed on the cluster \emph{Wheeler},
supported by the Sherman Fairchild Foundation and Caltech, and on
supercomputers of the NSF XSEDE network under  allocation
TG-PHY100033 and TG-PHY090003.  This paper has Yukawa Institute report
number YITP-17-40.

\vspace*{-0.3em}

\section{Supplemental Material}

\section{Analysis Details}

\textbf{\textit{Gravitational Wave Mismatch Calculation.}} The mismatch between
two observed waveforms $h^1(t)$ and $h^2(t)$ is defined as one minus the
maximum overlap $\mathcal{O}(h^1,h^2)$,
\begin{equation}
  \mathcal{M}(h^1,h^2) = 1 - \max_{ \{\chi_i\}}\mathcal{O}(h^1,h^2)\,,
\end{equation}
where the overlap is given by
\begin{equation}
\mathcal{O}(h^1,h^2) = \frac{\langle h^1|h^2\rangle}{\sqrt{ \langle h^1|h^1 \rangle \langle h^2|h^2 \rangle}}\,\,.
\label{eq:overlap}
\end{equation}
Here, $\langle \cdot | \cdot \rangle$ is a detector-noise weighted
inner product and optimization is carried out over a set $\{\chi_i\}$
of parameters impacting the overlap (e.g., shifts in waveform phases,
polarization angles, arrival times) \cite{damour:98}.

In the simplest case, we can choose $\langle \cdot | \cdot \rangle$ as
the frequency-domain noise weighted inner product \cite{cutler:94},
\begin{equation}
  \langle a|b \rangle_f = 4 \mathrm{Re} \int_0^\infty \frac{\tilde{a}(f) \tilde{b}^*(f)}{S_n(f)} df\,.
  \label{eq:scalar}
\end{equation}
Here, $S_n(f)$ is the detector noise power spectral density and
$\tilde{a}(f)$ is the Fourier transform of $a(t)$.

The real gravitational wave signal $h(t)$ observed by a single detector is given by
\begin{equation}
h(t) = F_+ h_+ + F_\times h_\times\,\,,
\label{eq:antenna}
\end{equation}
where $F_+$ and $F_\times$ are the detector antenna pattern functions that
depend on the sky location of the source and polarization basis (see,
e.g., \cite{harry:11}).

We now consider two scenarios: (1) A best case in which both $h_+$ and
$h_\times$ are measured by two optimally oriented GW detectors at
Advanced LIGO design sensitivity (``ZDHP'' for zero-detuning,
high-power \cite{LIGO-sens-2010}). (2) The realistic scenario of the
two Advanced LIGO interferometers with the sensitivity at the time of
GW150914.

For both cases, we need the two-detector inner product for two
detectors $\alpha$ and $\beta$, which is defined \cite{harry:11} as
the sum of the single-detector contributions,
\begin{equation}
  \langle h^1|h^2 \rangle_\mathrm{2 det} = \langle h^{1,\alpha}|h^{2,\alpha} \rangle_s + \langle h^{1,\beta}|h^{2,\beta} \rangle_s\,.
\label{eq:twodetector}
\end{equation}
Here, $h^{1,\alpha}$ is waveform 1 as seen by detector $\alpha$
through Eq.~\ref{eq:antenna} and so forth. The single-detector inner
product $\langle \cdot, \cdot \rangle_s$ used is that given by
Eq.~\ref{eq:scalar} with the exception that we integrate over some
frequency interval defined by $[f_\mathrm{min},f_\mathrm{max}]$. In practice, we obtain the necessary Fourier transforms by using the
Fast Fourier Transform algorithm after tapering the ends of the time
domain signal and padding with zeros for all waveforms to have the
same length in the time domain.

For scenario (1), we follow \cite{blackman:17} and define an optimal
two-detector $\mathcal{O}_\mathrm{opt}$ overlap by choosing detectors
oriented so that one detector is maximally sensitive to $h_+$ (and
insensitive to $h_\times$) while the opposite is true for the other
detector. We then have
\begin{equation}
\langle h^1|h^2 \rangle_\mathrm{opt} = \langle h_+^1| h_+^2 \rangle_s + \langle h_\times^1| h_\times^2 \rangle_s\,,
\end{equation}
with $S_n(|f|)$ in Eq.~\ref{eq:scalar} chosen as the Advanced LIGO
ZDHP noise power spectral density.  $\mathcal{O}_\mathrm{opt}$ is then
given by Eq.~\ref{eq:overlap} with $\langle \cdot | \cdot
\rangle_\mathrm{opt}$ and the mismatch is obtained as
$\mathcal{M}_\mathrm{ZDHP} = 1 - \max \mathcal{O}_\mathrm{opt}$.  We
optimize over time shifts and polarization angle shifts of the
waveforms. Since we consider only the $(2,2)$ GW mode, we simply
assume a face-on direction of GW propagation, and orbital phase shifts
are identical to polarization phase shifts.  See \cite{blackman:17}
for further details.
  
For scenario (2), we use the inner product of
Eq.~\ref{eq:twodetector} with the Advanced LIGO Hanford and Livingston
antenna patterns \cite{ligo-antenna} for GW150914 and the parameters
given in \cite{ligosys:17}. We employ the actual Hanford and
Livingston noise power spectral densities at the time of GW150914
provided at \url{https://losc.ligo.org/events/GW150914/}. We obtain
$\mathcal{M}_\mathrm{GW150914} = 1 - \max
\mathcal{O}_\mathrm{GW150914}$ for the $(2,2)$ GW mode by optimizing
over time shifts, polarization angle shifts, and orbital phase shifts.
We neglect contributions from other GW modes.

\textbf{\textit{Reduction in Log-Likelihood due to Mismatch.}} In GW parameter
estimation, the posterior probability of a BBH parameter vector
$\vec{\vartheta}$ is determined from the prior and likelihood. The GW
likelihood function (e.g., \cite{LIGO_properties:2016}) is given by

\begin{equation}
\label{loglikelihood}
\mathcal{L}(d|\vec{\vartheta}) \propto \, \mathrm{exp} \left[ -\frac{1}{2} \left \langle h^\mathrm{M}(\vec{\vartheta})-d | h^\mathrm{M}(\vec{\vartheta})-d  \right \rangle \right] \, . 
\end{equation}
Here, $d = h^\mathrm{GR} + n$ is the data observed in the detectors consisting of the GR signal (we use ``GR'' as a synonym for
``true'') and detector noise $n$. $h^\mathrm{M}$ is the template waveform
generated by some waveform model.

The log-likelihood is then
\begin{equation}
\begin{split}
\log \mathcal{L} = \mathrm{C} -\left[ \frac{1}{2} \left \langle h^\mathrm{M}|h^\mathrm{M} \right \rangle + \frac{1}{2} \langle h^{\mathrm{GR}}|h^{\mathrm{GR}}\rangle  \right. \\ \left. + \frac{1}{2} \langle n|n \rangle - \langle n | h^\mathrm{M}-h^\mathrm{GR}\rangle - \langle h^\mathrm{M} |h^{\mathrm{GR}} \rangle \right]  \, ,
\end{split}
\end{equation}
where $C$ is a constant of proportionality.

Suppose that $h^\mathrm{M}$ is different from the true signal, 
\begin{equation}
h^\mathrm{M} = (1+\epsilon_1)h^\mathrm{GR} + \epsilon_2 h^\bot \, ,
\end{equation}
where $\langle h^\bot | h^\mathrm{GR}\rangle = 0$. Here $\epsilon_1$
and $\epsilon_2$ are numbers and we consider the limit
$\epsilon_{1,2} \ll 1$. Any $h^\mathrm{M}$ can be decomposed in this way.
The log-likelihood becomes 
\begin{equation}
\begin{split}
\log \mathcal{L} = \log \mathcal{L}_0 - \frac{1}{2}\epsilon_1^2 \langle h^{\mathrm{GR}}|h^{\mathrm{GR}}\rangle - \frac{1}{2} \epsilon^2_2 \langle h^{\mathrm{\bot}}|h^{\mathrm{\bot}}\rangle \\ +\epsilon_1 \langle n|h^{\mathrm{GR}}\rangle + \epsilon_2 \langle n|h^\bot\rangle \, ,
\end{split}
\end{equation}
where $\log \mathcal{L}_0$ is the log-likelihood when $h^{M}=h^{\mathrm{GR}}$. The
expected reduction in the log-likelihood is then
\begin{equation}
\mathrm{E}[\delta \log\mathcal{L} ] = \frac{1}{2}\epsilon_1^2\langle h^\mathrm{GR}|h^\mathrm{GR}\rangle + \frac{1}{2}\epsilon_2^2\langle h^\bot|h^\bot\rangle \, .
\end{equation}

We now allow a small bias in the distance to the source by rescaling
$h^\mathrm{M}$ by $(1+\epsilon_1)^{-1}$ with which we obtain the convenient
expression
\begin{equation}
\mathrm{E}[\delta \log \mathcal{L} ] =  \frac{1}{2} \epsilon_2^2 \langle h^\bot | h^\bot \rangle + \mathcal{O}(\epsilon^3) \, .
\end{equation}

The mismatch between $h^{\mathrm{GR}}$ and $h^\mathrm{M}$ is
\begin{align}
\mathcal{M}(h^{\mathrm{GR}},h^\mathrm{M}) &= 1-  \frac{ \langle h^\mathrm{GR}|h^\mathrm{M}\rangle}{\sqrt{\langle h^\mathrm{GR}|h^\mathrm{GR}\rangle \langle h^\mathrm{M}|h^\mathrm{M} \rangle }} \\
	&= \frac{1}{2}\epsilon_2 \frac{\langle h^\bot|h^\bot\rangle}{\langle h^{\mathrm{GR}}|h^{\mathrm{GR}}\rangle} + \mathcal{O}(\epsilon^3) \, ,
\end{align}
where optimization over phase shifts, time shifts, etc. is implicit.

The signal-to-noise ratio $\varrho$ is given by $\varrho^2 = \langle
h^\mathrm{GR}|h^\mathrm{GR}\rangle$. With this, we find
\begin{equation}
\mathrm{E}[\delta \log\mathcal{L}] \approx \varrho^2 \mathcal{M}\,\,.
\end{equation}

The posterior probability will be affected by a factor of Euler's
number $e$ when $\delta \log\mathcal{L} = 1$, which can be considered
a mild observational inconsistency. Hence, the mismatch $\mathcal{M}$
will begin to have an effect on GW data analysis when
\begin{equation}
\mathcal{M} \gtrsim \frac{1}{\varrho^2} \, .
\end{equation}

\section{Numerical Convergence}

We carry out additional simulations at coarse-grid resolutions $\Delta
x_1=1.00 \, M$ and $\Delta x_3=1.60 \, M$, in addition to our
standard-resolution simulations of $\Delta x_2=1.25 \, M$. For our
convergence analysis, we choose the vacuum (G0) and the highest
density (G4) as two extremes of the simulations we carry out. We focus
our analysis on the gravitational waveforms since these are the most
important output of our simulations.

\begin{figure}[t]
  \hspace{-0.26cm}\includegraphics[width=\linewidth]{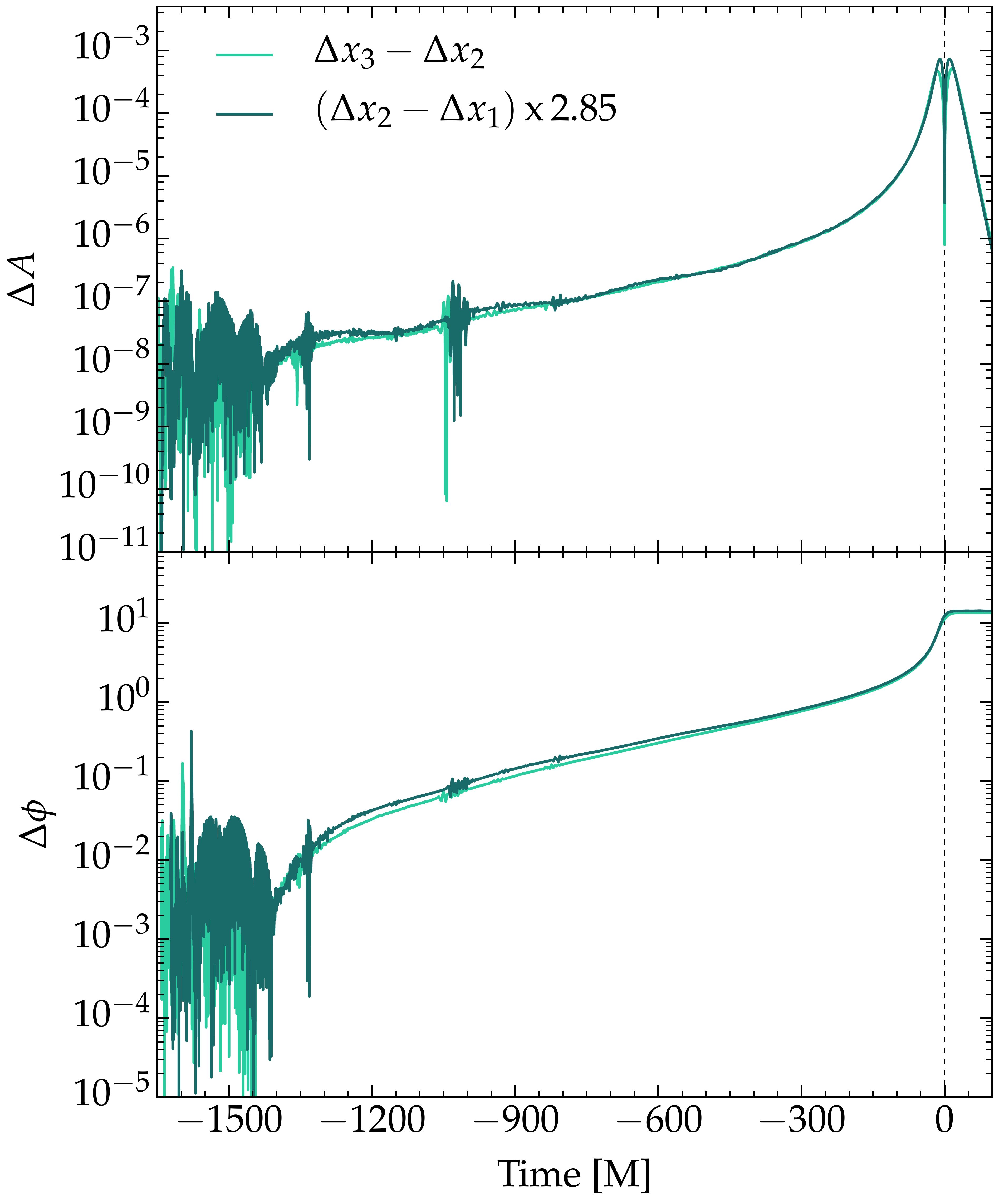}
  \caption{Fourth-order convergence for the G0 vacuum simulation. The dashed line at $0 \,M$ corresponds to merger, which we define as the maximum of the $h_{22}$ amplitude, and the time is given relative to merger. \textit{Top:} Amplitude differences between our lowest $(\Delta x_3)$, standard $(\Delta x_2)$, and highest-resolution $(\Delta x_1)$ simulations. We scale the differences using the self-convergence factor $1/Q_s=2.85$ corresponding to fourth-order convergence for this choice of resolutions. \textit{Bottom:} Phase angle differences also exhibiting fourth-order convergence.}
  \label{fig:vac_convergence}
\end{figure}

In Fig.~\ref{fig:vac_convergence}, we show numerical convergence in
the Newman-Penrose scalar $\psi_4$ between the different resolutions
for the G0 vacuum simulation. We consider phase and amplitude
differences separately. The amplitude is defined as
\begin{equation}
A(t) = \sqrt{\mathrm{Re}[\psi_{4}(t)]^2 + \mathrm{Im}[\psi_{4}(t)]^2}\, ,
\end{equation}
while the phase is defined as 
\begin{equation}
\phi(t) = \mathrm{tan}^{-1}\left( \frac{\mathrm{Im}[\psi_{4}(t)]}{\mathrm{Re}[\psi_{4}(t)]} \right)\, ,
\end{equation}
where $\mathrm{Re}[\psi_4]$ and $\mathrm{Im}[\psi_4]$ are the real and imaginary parts of $\psi_4$, respectively. Our numerical scheme is fourth-order, hence, we expect fourth-order convergence
and a self-convergence factor of
\begin{equation}
Q_s = \frac{\Delta x_3^n-\Delta x_2^n}{\Delta x_2^n-\Delta x_1^n} = 0.3505\,,
\end{equation}
where $n$ is the order of convergence. In
Fig.~\ref{fig:vac_convergence}, we rescale the differences between
highest resolution and second-highest (i.e.\ standard) resolution by
$1/Q_s$. These rescaled curves lie essentially on top of the curves
for the differences between second-highest and lowest resolution,
demonstrating approximate fourth-order convergence.

\begin{figure}[t]
	
  \hspace{-0.26cm}\includegraphics[width=1.0286\linewidth]{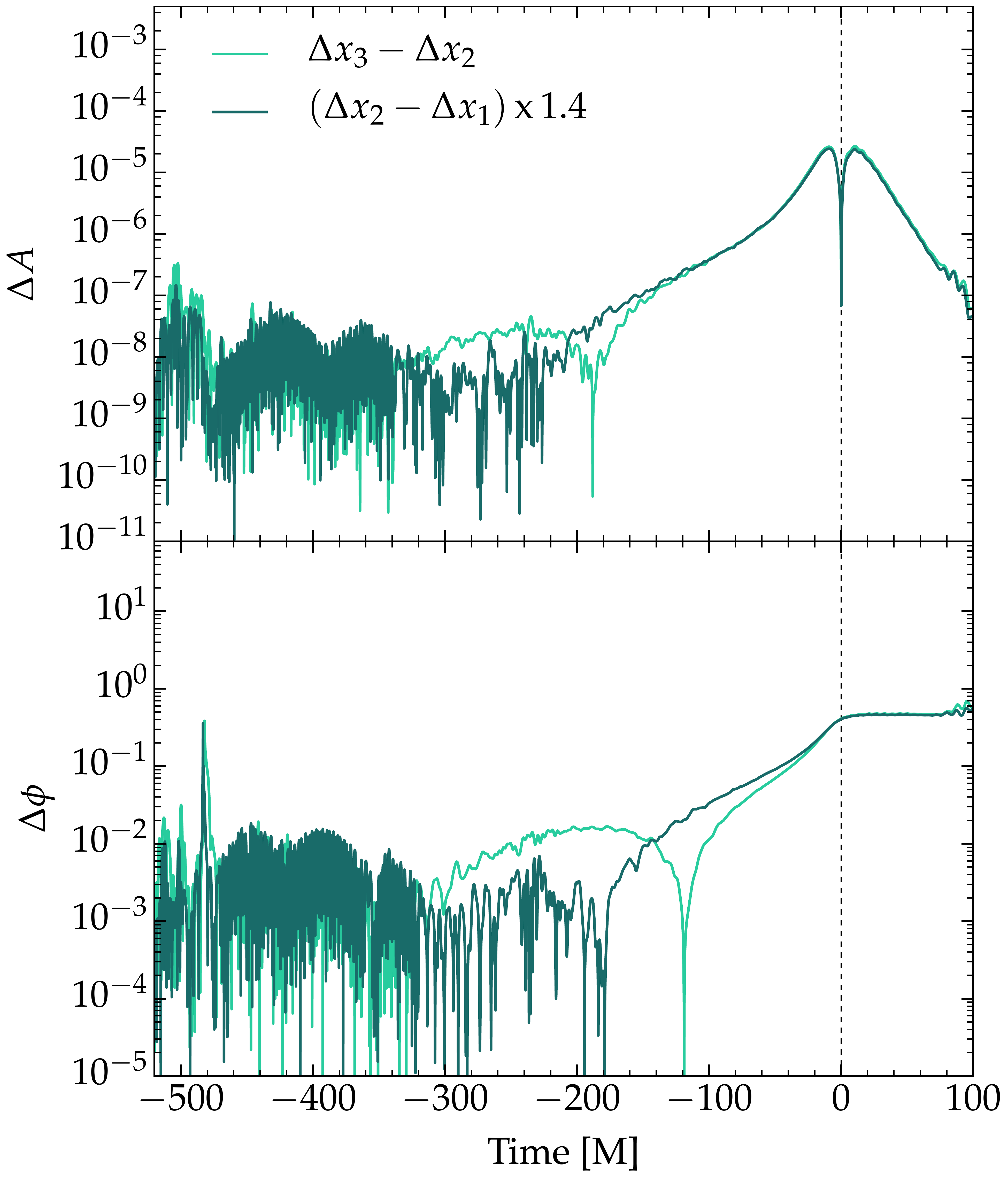}
  \caption{Convergence of the G4 simulation. The waveforms are aligned
    at merger, as defined in Figure 1, and all times are given relative to merger. The merger
    time is $0\, M$, marked with a dashed vertical line. \textit{Top:}
    Difference in waveform amplitude. We scale the difference between
    $\Delta x_2$ and $\Delta x_1$ by a self-convergence factor of
    $1/Q_s=1.4$, corresponding to first-order convergence. These are
    the simulations with the highest gas density and the evolution
    shows steep density gradients and shocks. Hence, we expect
    first-order convergence. \textit{Bottom:} Phase angle differences
    between the different resolution pairs, also exhibiting
    approximate first-order convergence.}
  \label{fig:neg7_convergence}
\end{figure}

In Fig.~\ref{fig:neg7_convergence} we perform the same analysis for
the highest-density simulation G4. In this case, the hydrodynamics
plays an important role in driving the coalescence. If our
finite-volume implementation dominates the numerical error, we expect
second-order convergence when the flow is smooth. However, soon after
the start of the simulation, steep density gradients and shocks
develop for which our numerical scheme (as any high-resolution shock
capturing scheme) is only first-order convergent. Hence, we can only
expect first-order convergence. We compute a first-order
self-convergence factor $Q_s = 0.7143$, with
$1/Q_s=2.85$. Figure~\ref{fig:neg7_convergence} shows that we obtain
roughly first-order convergence in GW amplitude and phase.

In order to clarify how numerical resolution effects the main results of our paper, we have calculate mismatches between various resolutions for the G0 and G4 cases. For the G0 case we find the mismatches to be $1.6 \times 10^{-3}$ between high and medium resolution, $2.9\times 10^{-3}$ between medium and low resolution, and $3.5\times 10^{-3}$ between high and low resolution. For the G4 case the mismatches are $3.5 \times 10^{-5}$ between high and medium resolution, $1.0 \times 10^{-4}$ between medium and low resolution, and $2.3 \times 10^{-4}$ between high and low resolution. Comparing these results with the mismatches listed in Table 1 (in the main paper), we conclude that our main conclusions are independent of numerical resolution.

\end{document}